\documentclass[11pt,letterpaper]{article}

\usepackage{graphicx} 
\usepackage[T1]{fontenc}
\usepackage[utf8]{inputenc}
\usepackage{lmodern}
\usepackage{upgreek}
\usepackage{textcomp}
\usepackage{amsmath,amsfonts,amssymb,bm}
\usepackage{latexsym}
\usepackage{amsthm}
\usepackage{mathrsfs}
\usepackage[english]{babel}
\usepackage{indentfirst,setspace,fancyhdr}
\usepackage[outercaption]{sidecap}
\usepackage[font=footnotesize]{caption}
\usepackage{geometry}
\usepackage{epstopdf}
\usepackage{color}
\usepackage[none]{hyphenat}
\usepackage[toc,page]{appendix}
\usepackage{multicol}
\usepackage{colortbl}
\usepackage{gensymb}
\usepackage{multirow}
\usepackage{array}
\usepackage{pstricks}
\usepackage{wrapfig}
\usepackage{subcaption}
\usepackage{fancyref}
\usepackage{pdfpages}
\usepackage{mathtools}
\usepackage{booktabs} 
\usepackage{quoting}
\usepackage{longtable}
\usepackage{cancel}
\usepackage{makecell}
\usepackage{tikz,stackengine}
\usetikzlibrary{matrix,calc}
\usepackage{mathtools}
\usepackage{blkarray}
\usetikzlibrary{decorations.pathreplacing}

\usepackage{pdflscape}
\usepackage{chngcntr}

\makeatletter
\renewcommand\@biblabel[1]{#1.} 
\makeatother
\allowdisplaybreaks

\emergencystretch=\maxdimen
\usepackage{authblk}

\usepackage{mathtools}
\usepackage{xcolor}
\usepackage{comment}
\usepackage{csquotes}
\usepackage[backend=biber, sorting=none]{biblatex}
\usepackage{lineno}

\makeatletter
\renewcommand\AB@maketitle{%
  \vskip -2em            
  {\parindent \z@ \centering
    \textbf{\@title} \par}%
  \vskip 1em
  {\centering \@author \par}
  \vskip 1em
  {\centering \@date \par}%
  \vskip 1em}
\makeatother

\usepackage[backgroundcolor=yellow!30,textsize=scriptsize]{todonotes}

\begin{document}

\onehalfspacing
\setlength\parindent{0pt}

\singlespacing 
\title{\large Exploring the conditions for sustainability with open-ended innovation}

\author{Debora Princepe$^{1,2,\spadesuit,*}$, Cristobal Quiñinao$^{3,\spadesuit}$, Cristina D\'iaz Faloh$^{4}$, \\ Pablo A. Marquet$^{3,5,6,7,8,*}$, and Matteo Marsili$^{1,2}$}

\affil{\small $^1$Quantitative Life Sciences Section, The Abdus Salam International Centre for Theoretical Physics (ICTP), Strada Costiera, 11 - 34151 - Trieste, Italy}
\affil{\small $^2$The Laboratory for Quantitative Sustainability, Viale Miramare 24/4 - 34151 - Trieste, Italy}

\affil{\small $^3$Facultad de Ciencias Biol\'ogicas - Pontificia Universidad Cat\'olica de Chile - 8331150 - Santiago, Chile}
\affil{\small $^4$Facultad de Física, Universidad de La Habana - 10400 - La Habana, Cuba}
\affil{\small $^5$Santa Fe Institute, Santa Fe - NM 87501 - USA}

\affil{\small $^6$Centro de Modelamiento Matemático (CMM), Universidad de Chile - IRL 2807 CNRS Beauchef 851 - 8370456 - Santiago, Chile}

\affil{\small $^7$Centro de Cambio Global UC, Avda. Vicuña Mackenna 4860 - 7820436 - Santiago,  Chile}

\affil{\small$^8$Instituto de Sistemas Complejos de Valpara\'iso (ISCV), Subida Artillería 470 - 2340000 - Valparaíso, Chile}

\affil{\small \textsuperscript{$*$} To whom correspondence should be addressed: deborapr@gmail.com, pmarquet@uc.cl}

\affil{\small \textsuperscript{$\spadesuit$} These authors contribute equally to this work}

\date{}

\maketitle

\begin{abstract}
Can sustained, open-ended socio-technical change preserve natural resources on a finite planet? 
We address this question using a stylized model in which an expanding repertoire of activities exploits a finite stock of environmental resources to drive population growth. 
In this model, an innovation event corresponds to a random draw of an activity characterized by a given  environmental impact and impact on population growth. 
An innovation successfully invades only if it satisfies constraints imposed by the current states of the environment and the population. 
We find that open-ended innovation leads to three possible outcomes: {\em i)} a doomsday scenario, with a finite time singularity in population growth and environmental collapse; {\em ii)} an equilibrium characterized by innovation-limited population growth, exhausted natural resources, and unfettered innovation; and {\em iii)} a sustainable regime characterized by Schumpeterian dynamics, where new ``greener'' activities displace older ones, preserving the environment while global population eventually stabilizes. 
The transition between these regimes is driven not by the intrinsic environmental impact of innovations but by how they interact with the population. 
These results align with the demographic transition observed in many countries, where high labor productivity tends to stabilize population growth, whereas population grows unbounded in economies with low labor productivity. 
By capturing the subtle feedbacks between socio-technical change, demography, and sustainability, our model provides a rational framework for understanding the determinants of the environmental Kuznets curve.
\end{abstract}

%

\section{Introduction}

The quantitative laws governing population growth have been discussed at least since Malthus, who pointed out their relation to natural resources~\cite{Malthus1803}. Much later, von Foerster {\it et al.}~\cite{von1960doomsday} showed that two millennia of data could be well described by a model leading the population to diverge at a finite time.  The doomsday was calculated to be November 26th 2026, and this behavior was traced back to superlinear effects induced by technological innovation. 

The importance of accounting for environmental constraints has been emphasized in landmark works such as Boulding's metaphor of ``Spaceship Earth'' \cite{boulding1966economics}, later expanded to include social well-being~\cite{raworth2017doughnut}, the ``Limits to Growth'' \cite{meadows1972limits}, and more recently the conceptual framework of planetary boundaries \cite{rockstrom2009planetary,steffen2015planetary,rockstrom2023safe}. Although some of these early works were criticized for neglecting market mechanisms and technological progress~\cite{solow1973end}, they stimulated a substantial bod
f research in economics aimed at integrating environmental impacts into economic analysis \cite{brock2013polluted,brander1998simple,brock2010green,arrow2012sustainability,dasgupta2021economics}. 

Innovation has been regarded both as a driver of environmental degradation and as a potential solution~\cite{boserup1981}. Yet, as observed by Jevons~\cite{alcott2005jevons}, even innovations that relieve environmental pressure through more efficient use of natural resources can lead to increased aggregate consumption and resource extraction. Regarding impacts on the population, innovations have generally improved productivity and living standards (yet see~\cite{mokyr2015history}), driving the {\it demographic transition} from high-fertility, low-income societies to low-fertility, high-income ones. Economic theories~\cite{galor2000population} attribute this transition to fertility choices that prioritize {quality over quantity}~\cite{barro1989fertility} in a world where technological progress favors a highly skilled labor force. 
This suggests that demographic growth is limited not by resource scarcity, as argued by Malthus, but by technological progress, as also argued by others~\cite{boserup1981,weinberger2017innovation,efferson2024our}, though under different premises.

This paper addresses whether open-ended innovation \textit{per se} is sustainable on a finite planet. 
Sustainability, insofar as it relates the needs of the current generation to the capacity of future generations to meet theirs, is an inherently anthropocentric notion whereby nature is at the service of the economy. Hence, much of the literature in economics has defined sustainability in terms of non-declining inter-generational utility of consumption~\cite{aghion1997endogenous}, or inclusive wealth~\cite{arrow2012sustainability,dasgupta2021economics}. In this view, a future in which natural resources are fully  depleted may be sustainable, in a weak sense~\cite{neumayer2013weak}, as long as these can be replaced by produced capital, whereas the preservation of natural resources -- i.e., sustainability in the strong sense~\cite{neumayer2013weak} -- is necessary only when critical  resources are non-substitutable. This anthropocentric perspective views nature and its resources as embedded within the economic system, rather than considering the human species as embedded in nature~\cite{daly1997beyond}. 

\begin{figure}[t]
    \centering
    \includegraphics[width=0.5\linewidth]{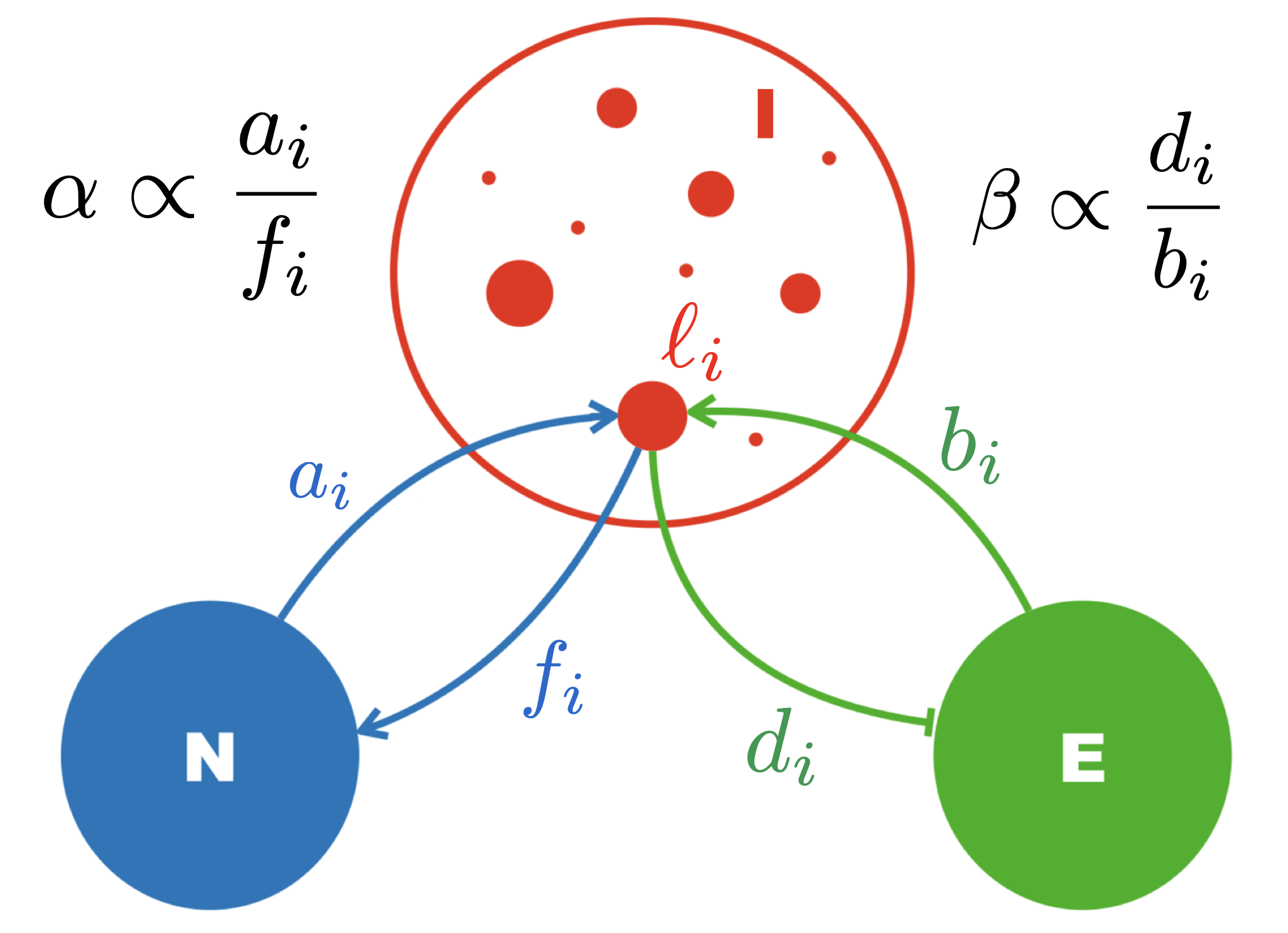}
    \caption{Illustrative sketch of the interaction between the sector of technologies ($I$), the environment ($E$), and the population ($N$). Arrows denote the interactions among the three components with parameters drawn independently at random for each technology $i = 1, \ldots, I$. The parameter $\alpha$ captures the interaction of technologies with the population, while $\beta$ quantifies the environmental impact of technologies.  }
    \label{fig:model}
\end{figure}

Here, we take a different perspective, modeling the entire system as an ecology where the population and the environment interact through technologies. 
Since our conclusions do not derive from the maximization of a welfare function~\cite{brander1998simple,dasgupta2021economics,barro2025economic,arrow2013sustainability}, we appeal to a basic notion of sustainability defined by the persistence of the population, of the environment, or of both, rather than in terms of non-decreasing well-being.
We take technological progress for granted, without discussing the conditions under which it may be promoted endogenously~\cite{romer1990endogenous,aghion1997endogenous}. We employ the term ‘technology’ in a broad sense to denote any activity that requires input from the population (labor) and the environment (natural resources), while affecting both of them (as illustrated in Fig. ~\ref{fig:model}). 
Our approach captures, albeit in a stylized form, the evolutionary nature of innovation and its capacity for unbounded increase in complexity~\cite{ruiz2008enabling,adams2017formal,corominas2018zipf}. Such open-ended dynamics are observed from cells to societies~\cite{hochberg2017innovation} and are intrinsic to cultural evolution~\cite{morgan2025human}. In economics, the idea of an open-ended innovation process was articulated by Schumpeter~\cite{schumpeter1942capitalism}, who argued that capitalist economies advance through long waves of growth fostered by key innovations 
driving older ones to extinction. 

We find that our model allows for both sustainable outcomes and an unsustainable ``doomsday'' scenario, where the population diverges in finite time~\cite{von1960doomsday}. 
Sustainable outcomes correspond to two types of equilibria separated by a phase transition. In the first, environmental resources vanish while the population grows as new technologies are introduced, which is consistent with a weak notion of sustainability~\cite{neumayer2013weak}. In the second, resources are preserved and the population saturates as new technologies displace less environmentally friendly incumbent ones, mirroring Schumpeter's concept of creative destruction. 
Interestingly, the driver of the transition between these regimes is not the environmental impact of technologies, but their impact on the population. Specifically, sustainable futures that preserve the environment are attained under conditions reminiscent of the combination of low fertility and high labor productivity prevalent in many industrialized countries.
We delve more deeply into on the significance of these results in the Discussion.


In the following sections, we introduce the model and present numerical results that agree with our theoretical analysis. All details of the calculations are relegated to \textit{Materials and Methods} and Supplementary Information.

\section{The Model}
\label{sec:model}

The model is schematically illustrated in Figure ~\ref{fig:model}. 
It describes the interplay between a finite ecosystem, a population, and a set of technologies. These components are represented by three variables: the environment $E$, the population $N$, and the number of technologies $I$. 
Each technology $i=1,\ldots,I$ is characterized by the number of individuals $\ell_i$ employed in it, as well as by the parameters that govern its interactions with the population and the environment. Innovation events correspond to random draws of these parameters from a distribution. In this sense, our model is based on a genuine, although stylized, innovation process, similar to the one introduced in ~\cite{bardoscia2017statistical}.

Between successive innovation events, the three components evolve as follows:

\begin{description}
    \item[Technologies] 
    For each technology $i$, the number of involved individuals $\ell_i$ evolves according to
    \begin{equation}
    \label{eq:st}
    \tau_L\frac{d\ell_i}{d t}=\max\left\{a_i\,(N-L)-b_i\,(E_0-E)-c\,\ell_i,0\right\}
    \,,\qquad L=\sum_{i=1}^I \ell_i\,.
    \end{equation}
    In words, all forms of capital that enter into an activity ultimately originate from either the population or the environment; Eq.~(\ref{eq:st}) describes how these sources combine to determine $\ell_i$.
    The term $a_i(N-L)-b_i(E_0-E)$ details how inputs $N$ and $E$ are transformed into output, measured in population units, by technology $i$, in a way similar to a production function in economics~\cite{barro2025economic}. 
    It is an increasing function of the available population~\cite{Fiaschi2025} $N-L$ and the stock of environmental resources $E$. The parameter $b_i$ tunes the dependence of a productive activity on the availability of natural resources, while $a_i$ determines how an additional unit of the (unemployed) population transforms into output. Primitive technologies requiring intensive human labor are characterized by small values of $a_i$, whereas more advanced ones, such as those run by automated processes with little human supervision, correspond to high $a_i$ values. Hence $a_i$ can be considered as a proxy of labor productivity.
    Note that Eq.~(\ref{eq:st}) assumes the substitutability of natural resources with labor input, as a decrease $dE$ in the stock of environmental resources can be compensated by an increase $dN=\frac{b_i}{a_i}dE$ in the population. 
    Finally, $c$ sets the decay rate of technology $i$ arising from labor force turnover or depreciation. A technology $i$ is introduced with $\ell_i=0$; if the parameters $a_i$ and $b_i$ are such that $a_i\,(N-L)-b_i\,(E_0-E)\le 0$, it does not invade, i.e., $\ell_i$ remains zero. 
    Taking $E_0$ as the initial value ($t=I=0$) of $E$ implies that at the beginning of the dynamics, when the environment is pristine, any technology would be able to invade. 
    We define $I_{\rm act}$ as the number of active technologies, which are those with $\ell_i>0$.
    
    \item[The environment] follows a standard equation in natural resource economics~\cite{clark2010mathematical,barro2025economic} 
    \begin{equation}
    \label{eq:E}
         \tau_E\frac{dE}{dt}= E\left(1-\frac{E}{E_0}\right)-\sum_i e_i(E)\ell_i E        \,,
    \end{equation}
    where the first term describes logistic growth that, in the absence of technologies, restores the environment to its pristine value $E_0$, while the second term represents the harvest function. The latter assumes that the depletion of natural resources is proportional to current stock $E$ and the effort level $\ell_i$ employed by each technology, where $e_i(E)$ is the extraction rate. We shall discuss both the case where $e_i(E)=d_i$ is a constant, and the case $e_i(E)=d_i\frac{E}{E_0}$ of an extraction rate proportional to $E$,  in order to describe a scenario in which resource extraction becomes increasingly difficult as resources become scarce. 
    
    \item[Population] Demographic growth follows a standard Malthus dynamics
    \begin{equation}
    \label{eq:N}
         \tau_N\frac{d N}{dt}= N\left(1+\sum_{i=1}^I f_i \ell_i-\frac{N}{N_0}\right)
    \end{equation}
    where the baseline population fitness is enhanced by technological progress and $N_0$ represents the carrying capacity in the absence of technology. The parameter $f_i$ captures the contribution of technology $i$ to fertility or population fitness. Notably, technologies that generate environmental pollution may diminish population fitness, for instance, by reducing reproductive success or increasing mortality.
\end{description}

The parameters $\tau_L$, $\tau_E$ and $\tau_N$ define the characteristic time scales for the dynamics of the different components. Each technology $i$ is defined by a set of parameters $\vec x_i=(a_i,b_i,d_i,f_i)$, all strictly positive and drawn from a probability distribution $p(\vec x)$. Technological progress is modeled as a sequence of innovation events, represented by random draws from $p(\vec x)$. We describe socio-technical innovation as a Poisson process with a rate that may depend on the state variables $N,E$ and $\ell_i$, $i=1,\ldots, I$.

The initial conditions are $E(t=0)=E_0$, $N(t=0)=N_0$, corresponding to the stationary state of Eqs.~\eqref{eq:E},~\eqref{eq:N} without technologies ($I=0$). We analyze the dynamics in terms of \textit{innovation time} $I$, which counts the number of innovations introduced up to time $t$, assuming that following each innovation event the dynamical system reaches its asymptotic state before the next innovation is introduced.

As previously stated, any innovation successfully invades a pristine environment ($E=E_0$) without technologies ($I=0$), as $\frac{d\ell_1}{dt}>0$ for small $\ell_1>0$. However, as the environment is depleted, the successful adoption of a new innovation requires that $N-L$ exceeds a threshold proportional to  environmental degradation, $E_0-E$.

\section{Results}
\label{sec:results}

Although the model can be studied in its full complexity, it is helpful to consider it in a simplified setting that is more easily tractable while allowing for a crisp interpretation of the results. Hence, we focus on the case where {\em i)} $a_i$ is proportional to $f_i$, and {\em ii)} $d_i$ is proportional to $b_i$. This allows us to analyze the model in terms of the two parameters 
\begin{equation}
    \alpha=\frac{1}{c E_0}\frac{a_i}{f_i}\,,\qquad\beta={c E_0}\frac{d_i}{b_i},
    \label{eq:alphabetadef}
\end{equation}
which have easily interpretable meanings. Specifically,  $\alpha$ takes low values when fertility ($f_i$) is high and/or with low labor productivity (low $a_i$), whereas large $\alpha$ relates to low-fertility and/or to high labor productivity (high $a_i$). Fixing the ratio of $a_i$ and $f_i$, as in Eq.~(\ref{eq:alphabetadef}), amounts to assuming that the more a technology depends on the population, the more it contributes to its growth.

The second of Eqs.~(\ref{eq:alphabetadef}) assumes that the more a technology depends on the flux of environmental services (large $b_i$), the more it degrades the environment (large $d_i$), with $\beta$ tuning the proportionality. Therefore, a world with environmentally friendly technologies corresponds to small values of $\beta$, whereas scenarios where technologies have substantial impact on the environment are described by large values of $\beta$.  

These assumptions allow us to analyze the model in terms of the two parameters $\alpha$ and $\beta$, yielding easily interpretable results. With $\alpha,\beta$ and $c$ fixed, an innovation is defined by the draw of only two random parameters, $a_i$ and $b_i$, which we assume to be drawn independently from exponential distributions with means $\bar a$ and $\bar b$, respectively. This choice is consistent with assuming that the only known quantities for $a_i$ and $b_i$ are their expected values, as the exponential is the maximum entropy distribution of a positive random variable with a given expected value. 

Whether an innovation is adopted ($\ell_i>0$) or not ($\ell_i=0$) is determined by the threshold $\kappa$. 
Specifically, adopted innovations must satisfy the condition 
\begin{equation}
\label{eq:kappa}
b_i\le \kappa a_i\,,\; \text{with} \; \kappa=\frac{N-L}{E_0-E},    
\end{equation}
which depends on the prevailing state of the population and the environment. We focus on the discussion of the results in what follows, referring to Section \textit{Materials and Methods} and Supplementary Information (SI) for their detailed derivation.

\begin{figure}[t]
    \centering
    \includegraphics[width=\linewidth]{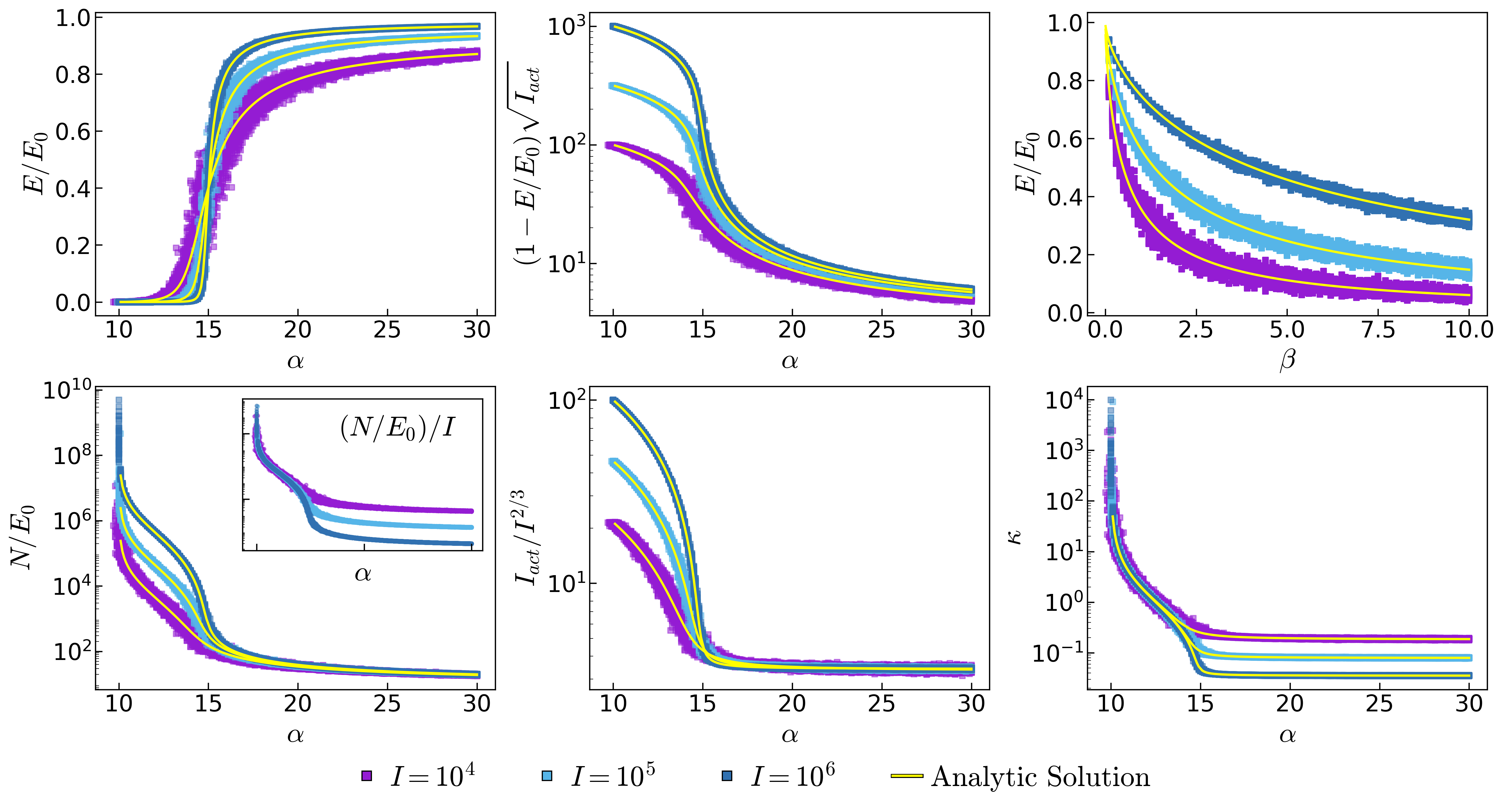}
    \caption{Results of numerical simulations of the linear model ($e_i\propto{E}$) with $I=10^4, 10^5$ and $10^6$ technologies, $N_0=100$ and $E_0=10$. $a_i$ and $b_i$ are independently drawn from exponential distributions with mean $c/2$ (see Materials and Methods).
    Top left: $E$ as a function of $\alpha$. Top middle: convergence of $E$ to $E_0$. Top right: $E$ as a function of $\beta$ (with $\alpha=20$). Bottom left: $N$ as a function of $\alpha$ (inset: $N/I$ as a function of $\alpha$). Bottom middle: $I_{\rm act}/I^{2/3}$ vs $\alpha$. Bottom right: $\kappa$ as a function of $\alpha$. 
    All quantities plotted as a function of $\alpha$ are obtained from the same set of simulations, with fixed $\beta = 0.1$.}
    \label{fig:results}
\end{figure}

Figure~\ref{fig:results} shows the results of the analytical calculations for different values of $I$ for the linear model ($e_i\propto E$), which are in excellent agreement with numerical simulations. The values of $E$, $N$, $I_{\rm act}$, and $\kappa$ are computed self-consistently by solving the equilibrium equations (see the \textit{Materials and Methods} and SI).
We distinguish three regimes, depending on the value of $\alpha$: 

\noindent
{\bf 1)}
For small $\alpha\le\alpha_u$ ($\alpha_u=10$ in Fig.~\ref{fig:results}), the dynamics is characterized by a finite-time singularity reminiscent of the \textit{doomsday} scenario~\cite{von1960doomsday}, where the environment is first completely depleted ($E=0$), and then population $N$ diverges. This occurs because the feedback of innovations on population growth -- the second term in the brackets of Eq.~(\ref{eq:N}) -- overcomes  the carrying capacity term. 
This mechanism is similar to the ``orgy of mutual benefaction'' effect discussed long ago by May~\cite{may1973stability}.
When $N$ grows very large, Eq.~(\ref{eq:N}) becomes $\frac{dN}{dt}\simeq qN^2$ for some constant $q$, leading to a finite-time singularity at a time inversely proportional to $q$ (see \textit{Materials and Methods} and SI).

\noindent
When $\alpha>\alpha_u$ at a fixed $I$, the dynamics asymptotically reaches a stationary state. In other words, for $\alpha>\alpha_u$, the asymptotic state of the system is limited by the available repertoire of technologies. We discuss this behavior assuming that  the system reaches its equilibrium for each value of $I$ before a new innovation ($I\to I+1$) is introduced. 

\noindent
{\bf 2)}
In the intermediate regime $\alpha_u<\alpha\le\alpha_c$ ($\alpha_c=15$ in Fig.~\ref{fig:results}), the environment is progressively depleted as more  innovations are introduced (as $E\propto I^{-1}$), as shown by comparing the three different curves of Fig.~\ref{fig:results} (Top Left). Correspondingly for $\alpha_u<\alpha\le\alpha_c$, population grows linearly with $I$ (Fig.~\ref{fig:results} Bottom Left). In this regime, open-ended innovation runs unfettered because $\kappa$ attains a finite limit as $I\to\infty$, and hence $I_{\rm act}$ grows linearly with $I$. This means that, in this hyper-innovative phase, old technologies are not replaced by new ones; instead, unfettered technological progress leads asymptotically to the complete exhaustion of environmental resources and sustained population growth $N\propto I$. 

\noindent
{\bf 3)}
When $\alpha\ge \alpha_c$ the system enters a sustainable phase where the environment converges to its pristine value and the population $N$ saturates at a finite value as  $I\to\infty$.  In this phase, the threshold $\kappa$ for the introduction of new technologies vanishes as $\kappa\propto I^{-1/3}$ as $I\to\infty$, while the number of active technologies grows as $I_{\rm act}\propto I^{2/3}$. As $\kappa$ decreases, 
the criteria for the successful introduction of new technologies become increasingly demanding. When adopted, innovations can cause a decrease in $\kappa$ that displaces existing technologies. This mechanism mirrors Schumpeter’s concept of creative destruction, so we refer to this as the \textit{Schumpeterian phase}.
In this regime, the environment approaches its pristine value as $E_0-E\propto I^{-1/3}$. At first glance, this result is surprising, as it implies that the sum $\sum_{i=1}^I d_i \ell_i$ in Eq.~(\ref{eq:E}) actually vanishes as $I\to\infty$, despite the number of terms growing with $I$. This is possible because, in this phase, the values of $d_i$ for successfully introduced technologies are bounded above by $\frac{\beta a_i}{c E_0}\kappa$, and $\kappa\sim I^{-1/3}$ as $I\to\infty$. In addition, the scale $\ell_i$ of adopted innovations also decreases as $I^{-2/3}$. Since the sum runs on $I_{\rm act}\sim I^{2/3}$ terms, the whole sum decreases as $I^{-1/3}$. Thus, in this regime, open-ended innovation ultimately benefits the environment.

These results refer to the model with a linear harvest rate $e_i(E)=d_i\frac{E}{E_0}$. In the case of a constant harvest rate $e_i(E)=d_i$, the intermediate regime does not exist, i.e. $\alpha_c=\alpha_u$, and the doomsday regime is followed directly by the Schumpeterian phase discussed above ($\alpha>\alpha_c$). 

Surprisingly, the preservation of the environment is not related to low environmental pressure (low $\beta$) as in the prevailing modeling approaches~\cite{brander1998simple,scheffer2001catastrophic,scheffer2009early}. Indeed, the parameter $\beta$ only induces smooth changes in the behavior of the model. 


\begin{figure}[t]
    \centering
    \includegraphics[width=\linewidth]{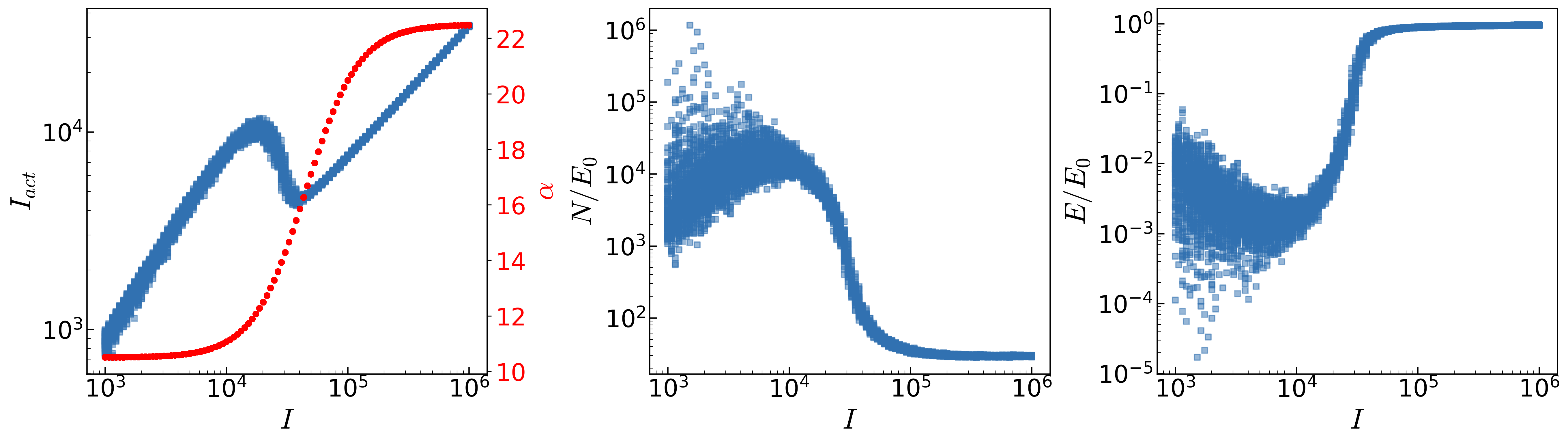}
    \caption{Demographic transition under increasing technological efficiency. Model simulations with a sigmoidal dependence of technological efficiency on $I$, $\alpha(I)=\alpha_0+x (I/I_0)^y/[1+(I/I_0)^y]$. Red points (secondary axis, leftmost panel) indicate $\alpha(I)$ for $x=12$, $y=2$, $I_0=4.5\times 10^4$, and $\alpha_0 = 10.5$.
    The subsequent recovery of the environment resembles the environmental Kuznets curve \cite{grossman1995economic}.}
    \label{fig:demo_trans}
\end{figure}

A developmental path where technological progress occurs along alongside increasing labor productivity and decreasing fertility rates, such as experienced by many economies, would be described by our model by an $\alpha(I)$ that increases with $I$. As shown in Figure~\ref{fig:demo_trans}, our model suggests that, provided the system avoids a finite-time singularity, this trajectory features a ``demographic transition'' to a regime where population saturates, as observed in several industrialized countries. Furthermore, along this path, the environment initially deteriorates but gradually recovers as the system enters the Schumpeterian phase. This behavior is reminiscent of the environmental Kuznets curve~\cite{grossman1995economic}.

\section{Discussion}

Our model reveals a demographic transition between two qualitatively distinct  regimes within the coupled population–environment–technology system. In one regime, technological innovation drives unbounded population growth and eventual environmental collapse. In the other, innovation follows a Schumpeterian process of creative destruction where new technologies replace existing ones. This competitive replacement leads to a sustainable equilibrium in which population stabilizes and environmental resources recover.

Contrary to the prevailing view~\cite{scheffer2001catastrophic,scheffer2009early}, the transition between these two regimes is not driven by the direct impacts of technologies on the environment (high $\beta$). Rather, it is governed by the parameter $\alpha$, which regulates the impact of technologies on population growth. High values of $\alpha$ correspond to technologies with high productivity and low labor requirements. In this regime ($\alpha>\alpha_c$), innovations compete for limited ecological and demographic resources, generating a Schumpeterian dynamics that restricts the expansion of technologies to increasingly environmentally friendly ones. 
As $\alpha$ decreases below $\alpha_c$, labor-intensive technologies proliferate and environmental depletion becomes unavoidable. When $\alpha$ falls below a second threshold $\alpha_u$, positive feedback between technological expansion and population growth drives the system toward a runaway trajectory, culminating in a finite-time singularity reminiscent of the ``doomsday'' scenario proposed by von Foerster {\em et al.}~\cite{von1960doomsday}. In the intermediate regime $\alpha_u<\alpha\le\alpha_c$, technological activities expand without bound while progressively replacing natural ecosystem services. This embodies a ``life on Mars'' outcome, where a self-sustaining technological ecosystem, an artificial biosphere, substitutes for the natural one. Despite the environmental collapse, this state may still satisfy weak notions of sustainability~\cite{neumayer2013weak} provided technological substitutes ensure non-decreasing levels of well-being.

The innovation dynamics generated by the model parallels patterns observed in biological evolution. In the Schumpeterian regime, innovation proceeds through competitive replacement analogous to periodic selection in microbial populations~\cite{atwood1951periodic}. Similar transitions toward diversification have been observed in long-term bacterial cultures under starvation stress, where genetically homogeneous populations give way to diverse communities sustained by cross-feeding and resource recycling~\cite{finkel1999evolution,keymer2012diversity,gerrish1998fate}. These analogies support the perspective that technological innovation and biological evolution share common dynamical features 
~\cite{nunez2014dynamics,weinberger2017innovation}.

From a socio-economic perspective, the model reproduces key features of the demographic transition observed during industrialization. 
Indeed, historical data indicate that global labor productivity has increased by more than an order of magnitude over the last century~\cite{rhode2005world}. In our framework, trajectories with expanding technological diversity alongside rising labor productivity, are those in which $\alpha(I)$ increases with $I$. Provided such trajectories avoid the singular regime, they naturally lead to population stabilization accompanied by gradual environmental recovery.
This pattern resembles the environmental Kuznets curve~\cite{grossman1995economic}, according to which environmental degradation initially increases during early development stages but declines once the economy matures. While its empirical validity remains debated~\cite{dasgupta2002confronting,stern2004rise}, our results suggest that such dynamics may arise generically from the coupled evolution of population, technology, and environmental resources, independent of specific microeconomic mechanisms.

These findings contribute to ongoing debates regarding the role of technological innovation in sustainability. Market-driven technological change has historically supported economic growth~\cite{galor2000population,aghion1997endogenous}, yet anthropogenic pressures are pushing several planetary processes toward critical boundaries~\cite{rockstrom2009planetary}. This tension has stimulated contrasting views: ecomodernist perspectives emphasize the potential of technological innovation to facilitate sustainable futures, by decoupling human activity from the environment~\cite{asafuAdjaye2015ecomodernist}, whereas other approaches highlight the need for institutional change and governance of global commons~\cite{heron2025anti,daly1997beyond,ostrom1990governing}. 

By abstracting from explicit economic mechanisms, our approach contributes to this debate by highlighting a key ingredient that institutional change in governance structures should target. Specifically, our results reveal that a fundamental driver of sustainability lies in the interaction between technological innovation and population dynamics. Sustainable outcomes arise in our model not because technologies reduce their direct environmental impacts (low $\beta$), but because they weaken the dependence of economic activity on labor input. Under these conditions (high $\alpha$), environmentally damaging activities may be gradually displaced by more sustainable ones. This mechanism effectively decouples population and environmental resources.

Therefore, our results suggest that policies that enhance economic security -- such as universal basic income~\cite{langridge2023ecological} -- could indirectly foster sustainability by lowering incentives for labor-intensive activities. In this view, sustainable management of the Earth system requires governance mechanisms capable of balancing human needs with planetary constraints~\cite{raworth2017doughnut,ostrom1990governing,Fanning2025}.

The simplicity of our framework entails important limitations. The model neglects institutional, economic, and political processes shaping technological change~\cite{weisz2015industrial}, as well as the dependence of technological systems on non-substitutable natural resources~\cite{graedel2015materials}. Moreover, we simplify the dynamics of the environment to that of a natural resource but we ignore the complex  nonlinear dependencies within natural ecosystems that could give rise to  thresholds and tipping points, as a consequence of global changes such as biodiversity loss, climate warming, and pollution ~\cite{rockstrom2009planetary,scheffer2001catastrophic}. We hope that the present work will stimulate further research exploring the complex interactions between technological innovation, demographic dynamics, and environmental sustainability.

\section{Materials and Methods}



We first rewrite Eqs.~\eqref{eq:st},~\eqref{eq:E} and \eqref{eq:N} in terms of dimensionless variables and parameters. To do this, we measure $N$, $\ell_i$ and $E$ in units of a reference biomass $E_0$. Therefore the parameters $a_i,b_i$ and $c$ are dimensionless, while $d_i$ and $f_i$ have dimensions of inverse biomass. 
We introduce the dimensionless variables $\eta$, $\lambda_i$, and $\epsilon$
\begin{equation}
    \label{eq:def:resc_vars}
    \eta=\frac{N}{E_0}\,,\qquad \lambda_i=\frac{\ell_i}{E_0}\,,\qquad\epsilon=\frac{E}{E_0}\,,
\end{equation}
and parameters $\theta_i$, $\gamma_i$, $\sigma_i$ and $\delta_i$
\begin{equation}
    \label{eq:def:resc_params}
    \theta_i=\frac{a_i}{c}\,,\qquad \gamma_i=\frac{b_i}{c}\,,\qquad\sigma_i={E_0}f_i\,,\qquad\delta_i=E_0 d_i\,.
\end{equation}
In terms of these quantities and the dimensionless time $s=t/\tau_L$, Eqs.~\eqref{eq:st},~\eqref{eq:E} and \eqref{eq:N} read
\begin{eqnarray}
\label{eq:lambda}
    \frac{d\lambda_i}{ds}&=&c\max\left\{\theta_i(\eta-\Lambda)-\gamma_i(1-\epsilon)-\lambda_i,0\right\}\,,\qquad \Lambda=\frac{L}{E_0}=\sum_{i=1}^I\lambda_i \\
\label{eq:eta}
    \frac{\tau_N}{\tau_L}\frac{d\eta}{ds}&=&\eta\left(1+\sum_{i=1}^I\sigma_i\lambda_i-\frac{\eta}{\eta_0}\right)\,,\qquad\eta_0=\frac{N_0}{E_0}\\
\label{eq:epsilon}
    \frac{\tau_E}{\tau_L}\frac{d\epsilon}{ds}&=&\epsilon(1-\epsilon)-\sum_{i=1}^I\delta_i \lambda_i\epsilon^{1+u}\,,\qquad u=0,1
\end{eqnarray}
where $u=0$ corresponds to a constant extraction rate, while $u=1$ describes extraction rates proportional to the fraction of available resources $\epsilon$. In terms of the dimensionless parameters, we have $\alpha=\theta_i/\sigma_i$ and $\beta=\delta_i/\gamma_i$. We assume that $\theta_i$ and $\gamma_i$ are independently drawn from exponential distributions with means $\bar\theta$ and $\bar\gamma$, respectively, and then take $\sigma_i=\theta_i/\alpha$ and $\delta_i=\beta\gamma_i$.

\subsection*{The innovation limited regime $\alpha>\alpha_u$}

Let us first discuss the regime where, with a finite number $I$ of technologies, the dynamics reaches a stationary state ($\alpha>\alpha_u$ ). We assume that the time for introduction of the next innovation is much longer that the time for the system to relax to the stationary state, which means that we shall study the system in innovation time $I$. Setting Eq.~(\ref{eq:lambda}) to zero yields 
\begin{equation}
    \lambda_i=\theta_i(\eta-\Lambda)-\gamma_i(1-\epsilon)
    \label{eq:lambda_eq}
\end{equation}
for all active technologies. Summing over all $i$ for which $\lambda_i>0$, we obtain
\begin{equation}
    \Lambda=\frac{\Theta\eta}{1+\Theta}-\frac{\Gamma(1-\epsilon)}{1+\Theta}\,,\qquad\Theta=\sum_{i:\lambda_i>0}\theta_i\,,~~\Gamma=\sum_{i:\lambda_i>0}\gamma_i\,.
\end{equation}
The condition for an innovation to be adopted now reads $\gamma_i\le\kappa\theta_i$, where the threshold 
\begin{equation}
\kappa=\frac{\eta-\Lambda}{1-\epsilon}=\frac{1}{1+\Theta}\left(\frac{\eta}{1-\epsilon}+\Gamma\right)    
\end{equation}
can be expressed in terms of the state variables $\eta$ and $\epsilon$ and the coefficients $\Theta$ and $\Gamma$ that depend on the active technologies. This in turn allows us to compute averages over the adopted technologies and to estimate the sums on $\sigma_i\lambda_i$ and $\delta_i\lambda_i$ appearing in Eqs.~\eqref{eq:eta},~\eqref{eq:epsilon} for $I\gg 1$. The stationary states of these equations ultimately yields closed equations for $\eta$ and $\epsilon$. 
These equations depend on coefficients such as $\Theta$ and $\Gamma$ that can be estimated self-consistently using the law of large numbers in the limit $I\to\infty$. 

\subsection*{The doomsday regime $\alpha\le \alpha_u$}

We further assume that $\tau_L\ll\tau_N$, so that $\lambda_i$ can be taken to be at its stationary state for all values of $\eta$, $\epsilon$, and $I$. Using the expressions above for $\lambda_i$ and $\Lambda$, and the calculations reported in the next subsection, the equation for $\eta$ reads
\begin{equation}
    \frac{\tau_N}{\tau_L}\frac{d\eta}{ds}=\eta[1+g\eta+h(1-\epsilon)]\,,
    \label{eq:dynsing}
\end{equation}
where $g=\frac{\Psi}{\alpha(1+\Theta)}-\eta_0^{-1}$ and $h=\left(\frac{\Gamma\Psi}{1+\Theta}-\Delta\right)/\alpha$, with
\begin{equation}
    \Psi=\sum_{i:\lambda_i>0}\theta_i^2\,,~~\Delta=\sum_{i:\lambda_i>0}\theta_i\gamma_i\,.
\end{equation}
When $g,h>0$, the solution of Eq.~(\ref{eq:dynsing}), in real time $t$, reads 
\begin{equation}
    \eta(t)=\frac{\eta_0}{\left(1+g\eta_0/h\right)e^{-ht/\tau_N}-\eta_0g/h},
\end{equation}
which has a finite time singularity at $t_c=\frac{\tau_N}{h}\log\left(1+\frac{h}{\eta_0 g}\right)$. The condition $g>0$ yields the condition that defines $\alpha_u=\frac{\eta_0\Psi}{1+\Theta}$. Since $\Theta$ and $\Psi$ depend on the innovations introduced up to $I$, the singularity may manifests when a specific innovation is introduced. Notice that for $I\to\infty$, $\alpha_u$ reaches a finite limit because of the law of large numbers. The singularity persists as long as $h>-g\eta_0$. For large $I$ this requires that the covariance between $\theta_i$ and $\gamma_i$ be smaller than the variance of $\theta_i$ divided by the mean of $\gamma_i$. In our setting that assumes that $\gamma_i$ and $\theta_i$ are independent, this condition is satisfied.

\subsection*{Equilibrium of the system}

We examine the equilibrium of the system, setting all derivatives of equations (\ref{eq:lambda}), (\ref{eq:eta}), and (\ref{eq:epsilon}) to zero. For $\lambda_i$, we get Eq. (\ref{eq:lambda_eq}).
To obtain the equivalent equation for the steady states of the population in terms of $\eta$, we write the equilibrium value for technology $i$ in terms of $\epsilon$ and $\eta$ alone,
\begin{equation}
    \lambda_i=\frac{\theta_i}{1+\Theta} \eta+\theta_i\left(\frac{\Gamma}{1+\Theta}-\frac{\gamma_i}{\theta_i}\right)(1-\epsilon),
\end{equation} 
and, assuming $\sigma_i=\theta_i/\alpha$, then we have
\begin{align*}
    \alpha\sum_{i}\sigma_i \lambda_i &=\sum_{i}\left(\frac{\theta_i^2}{1+\Theta}\eta+\theta_i^2\left(\frac{\Gamma}{1+\Theta}-\frac{\gamma_i}{\theta_i}\right)(1-\epsilon)\right) 
   \\
   &=\frac{\Psi}{1+\Theta}\eta + \left(\frac{\Psi\Gamma}{1+\Theta}-\Delta\right) (1-\epsilon).
\end{align*}
Using this relationship, we can rewrite the steady state for $\eta$ as
 \[\eta  = \eta_0 (1+\sum_{i=1}^I \sigma_i \lambda_i) = \eta_0+\frac{\eta_0}{\alpha}\frac{\Psi}{1+\Theta}\eta + \frac{\eta_0}{\alpha}\left(\frac{\Psi\Gamma}{1+\Theta}-\Delta\right) (1-\epsilon),
 \]
then
\begin{align}
    \eta  = \eta_0 \frac{1+\frac{\Sigma}{\alpha}(1-\epsilon)}{1-\eta_0\frac{\chi}{\alpha}},\qquad\text{with}\quad
\chi = \frac{\Psi}{1+\Theta},\quad \Sigma=\left(\frac{\Psi\Gamma}{1+\Theta}-\Delta\right).
\label{eq:eq_eta}
\end{align}

For the equation of the environmental variable $\epsilon$, we proceed similarly. If $\delta_i=\beta\gamma_i$, then 
\begin{align*}
   \frac{1}{\beta}\sum_{i}\delta_i \lambda_i &=\sum_i\left(\frac{\gamma_i\theta_i}{1+\Theta}\eta+\gamma_i\theta_i\left(\frac{\Gamma}{1+\Theta}-\frac{\gamma_i}{\theta_i}\right)(1-\epsilon)\right) 
   \\
   &=\frac{\Delta}{1+\Theta}\eta + \left(\frac{\Delta\Gamma}{1+\Theta}-\Phi\right) (1-\epsilon),
\end{align*}
with $\Phi=\sum_{i}\gamma_i^2$.  

We focus in the case of extraction rate proportional to $\epsilon$, $u=1$, which exhibits a richer dynamics. The $u=0$ case is addressed in the SI. 
The equilibrium for $\epsilon$ implies
\begin{equation*}
    1-\epsilon=\epsilon \sum_i \delta_i \lambda_i= \epsilon \beta \left( \frac{\Delta}{1+\Theta}\eta - \Omega (1-\epsilon)\right).
\end{equation*}
By using~\eqref{eq:eq_eta}, it follows that
    \begin{align*}
   \frac{1}{\epsilon} &=1+\beta\frac{ \Delta}{1+\Theta}\eta_0 \frac{1+\frac{\Sigma}{\alpha}(1-\epsilon)}{1-\eta_0\frac{\chi}{\alpha}} - \beta\Omega (1-\epsilon)
   \\
   &=1+\frac{\beta \Delta}{1+\Theta}\frac{\eta_0}{1-\eta_0\frac{\chi}{\alpha}}+ \beta (1-\epsilon)\left(\frac{\Sigma\Delta\eta_0}{\alpha(1+\Theta)(1-\eta_0\frac{\chi}{\alpha})}-\Omega\right)
   \\
   &=1+A+B(1-\epsilon),
  \end{align*}
  where
  \begin{equation}
    A=\frac{\beta\Delta\eta_0}{(1+\Theta)(1-\eta_0\frac{\chi}{\alpha})}\,,\qquad B=\beta\left[\frac{\Delta\Sigma \eta_0}{\alpha(1+\Theta)(1-\eta_0\frac{\chi}{\alpha})}-\Omega\right].
\end{equation}
The solution for $\epsilon$ is 
\begin{equation}
\label{eq:constN0}
    \epsilon=\frac{1+A+B\pm\sqrt{(1+A+B)^2-4B}}{2B}
\end{equation}
The coefficient $A$ diverges when $\eta_0\frac{\chi}{\alpha}\to 1$, which marks the instability transition at $\alpha_u=\eta_0 \chi$.

\subsection*{Algorithm for numerical simulations} 

Time is indexed by the sequence of innovations; specifically, at step $i$, the number of technologies introduced is $i$.
Each realization starts at  $i = 0$, with initial conditions $\epsilon(0) = 1$ and $\eta(0) = \eta_0$ ($\eta_0$ taken constant without loss of generality), and fixed model parameters $\alpha$ and $\beta$. The values adopted in the manuscript correspond to $\eta_0 = 10$.

Starting from a state with no active technologies ($I_{act} = 0$), we attempt to activate technologies sequentially.
At each step, we compute the updated equilibrium values of the environment and population, and check if the next technology satisfies the activation condition, i.e., whether it can successfully invade. The process stops as soon as this condition is no longer met.
Computationally, the procedure is as follows:
\begin{enumerate}
    \item Generate an ensemble of $I$ technologies $(\gamma_i, \theta_i)$, sampled independently from exponential distributions with means $\bar{\gamma}$ and $\bar{\theta}$, respectively. For the results presented, we set $\bar{\gamma} = \bar{\theta} = 0.5$ as a convenient parameter choice.
    
    \item Compute $\kappa_i = \gamma_i / \theta_i$, and sort the technologies in increasing order of $\kappa_i$.
    
    \item Loop through the sorted list, attempting to activate one technology at a time:
    \begin{itemize}
        \item * Increment the counter: $I_{act} \leftarrow I_{act} + 1$, and consider the technology  $\kappa_{I_{act}}$. Compute the accumulated variables $\Theta$, $\Gamma$, $\Delta$, $\Psi$, and $\Phi$, and determine the corresponding equilibrium values of $\epsilon$ and $\eta$;
        
        \item Compute the updated threshold $\kappa$, and check whether the next technology satisfies the activation condition given by Eq.~\ref{eq:kappa}, $\kappa_{I_{act} + 1} \leq \kappa$;
        
        \item If the condition is violated, stop the process and proceed to the next realization. Store the current values of the environment, population, and the last successfully activated $\kappa_{I_{act}}$. Otherwise, return to step *.
    \end{itemize}
\end{enumerate}

The algorithm yields, for each realization, the final number of active technologies $I_{act}$, the equilibrium values of $\epsilon$ and $\eta$, and the maximum accepted value of $\kappa_i$. We performed 100 realizations for each value of $\alpha$ or $\beta$. 


\section*{Acknowledgment}

We thank the participants of the TLQS Workshop on Limits to Collective Agency, held at the Abdus Salam ICTP (6-10 May 2024), for simulating and fruitful discussions.  We are particularly grateful to Erol Ak\c{c}ay, Davide Fiaschi, Jacopo Grilli, Thomas Hills, May Lim, Onofrio Mazzarisi and Mercedes Pascual for their valuable comments and insights. Funding from the National Institute of Oceanography and Applied Geophysics (OGS), within The Laboratory for Quantitative Sustainability, is gratefully acknowledged. P.A.M.  and C.Q. acknowledge support from Exploration Grant 13220168 from The Agencia Nacional de Investigación y Desarrollo (ANID). P.A.M. also acknowledges support from the ICTP through its Associates Programme, funded by the Simons Foundation through grant number 284558FY19, and from Grant FB210005, BASAL funds for Centers of Excellence from ANID-Chile.


\section*{Author contributions}
P. A. M. and M. M. designed the study; D. P., C. D. F. and M. M. run the simulations; D. P., C. Q., C. D. F., P. A. M. and M. M. performed the analytic calculations and analyzed the results; D. P., C. Q., P. A. M. and M. M. wrote the original draft.

\section*{Conflict of interest}
{The authors declare no conflict of interests.}

\section*{ Code availability }
The codes used in this study are available in the GitHub repository at \url{https://github.com/deborapr/evol_technologies}.

\printbibliography

@article{efferson2024our,
  title={Our fragile future under the cumulative cultural evolution of two technologies},
  author={Efferson, Charles and Richerson, Peter J and Weinberger, Vanessa P},
  journal={Philosophical Transactions of the Royal Society B},
  volume={379},
  number={1893},
  pages={20220257},
  year={2024},
  publisher={The Royal Society}
}

@misc{Fiaschi2025,
  author = "Davide Fiaschi",
  howpublished = "Private communication",
  year = {2025},
  note = "An alternative interpretation of the first term of Eq.~(1), in line with a Schumpeterian perspective, is that the expansion of a productive activity is driven by the demand. In this view, $\ell_i$ would be the consumption of the good produced by the $i^{\rm th}$ productive activity. Assuming that each individual demands one unit of good, the growth of $\ell_i$ would be proportional to the excess demand $N-L$."
}

@article{steffen2015planetary,
  title={Planetary boundaries: Guiding human development on a changing planet},
  author={Steffen, Will and Richardson, Katherine and Rockstr{\"o}m, Johan and Cornell, Sarah E and Fetzer, Ingo and Bennett, Elena M and Biggs, Reinette and Carpenter, Stephen R and De Vries, Wim and De Wit, Cynthia A and others},
  journal={science},
  volume={347},
  number={6223},
  pages={1259855},
  year={2015},
  publisher={American Association for the Advancement of Science}
}

@article{rockstrom2009planetary,
  title={Planetary boundaries: exploring the safe operating space for humanity},
  author={Rockstr{\"o}m, Johan and Steffen, Will and Noone, Kevin and Persson, {\AA}sa and Chapin III, F Stuart and Lambin, Eric and Lenton, Timothy M and Scheffer, Marten and Folke, Carl and Schellnhuber, Hans Joachim and others},
  journal={Ecology and society},
  volume={14},
  number={2},
  year={2009},
  publisher={JSTOR}
}

@article{finkel1999evolution,
  title={Evolution of microbial diversity during prolonged starvation},
  author={Finkel, Steven E and Kolter, Roberto},
  journal={Proceedings of the National Academy of Sciences},
  volume={96},
  number={7},
  pages={4023--4027},
  year={1999},
  publisher={The National Academy of Sciences}
}

@article{nunez2014dynamics,
  title={The dynamics of technological change under constraints: Adopters and resources},
  author={N{\'u}{\~n}ez-L{\'o}pez, M and Velasco-Hern{\'a}ndez, JX and Marquet, PA},
  journal={Discrete and Continuous Dynamical Systems-B},
  volume={19},
  number={10},
  pages={3299--3317},
  year={2014},
  publisher={Discrete and Continuous Dynamical Systems-B}
}

@article{keymer2012diversity,
  title={Diversity emerging: from competitive exclusion to neutral coexistence in ecosystems},
  author={Keymer, Juan E and Fuentes, Miguel A and Marquet, Pablo A},
  journal={Theoretical Ecology},
  volume={5},
  pages={457--463},
  year={2012},
  publisher={Springer}
}

@article{gerrish1998fate,
  title={The fate of competing beneficial mutations in an asexual population},
  author={Gerrish, Philip J and Lenski, Richard E},
  journal={Genetica},
  volume={102},
  pages={127--144},
  year={1998},
  publisher={Springer}
}

@article{atwood1951periodic,
  title={Periodic selection in Escherichia coli},
  author={Atwood, Kimball C and Schneider, Lillian K and Ryan, Francis J},
  journal={Proceedings of the National Academy of Sciences},
  volume={37},
  number={3},
  pages={146--155},
  year={1951}
}

@article{alcott2005jevons,
  title={Jevons' paradox},
  author={Alcott, Blake},
  journal={Ecological economics},
  volume={54},
  number={1},
  pages={9--21},
  year={2005},
  publisher={Elsevier}
}

@book{boserup1981,
  title={Population and Technological Change},
  author={Boserup, Ester},
  year={1981},
  publisher={University of Chicago Press, Chicago}
}

@book{meadows1972limits,
  title={The Limits to Growth},
  author={Meadows, Donella H and Meadows, Dennis L and Randers, J{\o}rgen and Behrens III, William W},
  year={1972},
  publisher={Universe Books, New York}
}

@book{dasgupta2021economics,
  title={The Economics of Biodiversity: The Dasgupta review},
  author={Dasgupta, Partha},
  year={2021},
  publisher={(London: HM Treasury}
}

@article{rockstrom2023safe,
  title={Safe and just Earth system boundaries},
  author={Rockstr{\"o}m, Johan and Gupta, Joyeeta and Qin, Dahe and Lade, Steven J and Abrams, Jesse F and Andersen, Lauren S and Armstrong McKay, David I and Bai, Xuemei and Bala, Govindasamy and Bunn, Stuart E and others},
  journal={Nature},
  volume={619},
  number={7968},
  pages={102--111},
  year={2023},
  publisher={Nature Publishing Group UK London}
}

@article{von1960doomsday,
  title={Doomsday: Friday, 13 November, AD 2026: At this date human population will approach infinity if it grows as it has grown in the last two millenia.},
  author={Von Foerster, Heinz and Mora, Patricia M and Amiot, Lawrence W},
  journal={Science},
  volume={132},
  number={3436},
  pages={1291--1295},
  year={1960},
  publisher={American Association for the Advancement of Science}
}

@incollection{boulding1966economics,
  title={The economics of the coming spaceship earth},
  author={Boulding, Kenneth E},
  booktitle={Environmental quality in a growing economy},
  editor={Jarrett, H.},
  pages={3--14},
  year={1966},
  publisher={Resources for the Future/John Hopkins University Press}
}

@article{brock2010green,
  title={The green Solow model},
  author={Brock, William A and Taylor, M Scott},
  journal={Journal of Economic Growth},
  volume={15},
  pages={127--153},
  year={2010},
  publisher={Springer}
}

@article{arrow2013sustainability,
  title={Sustainability and the measurement of wealth: further reflections},
  author={Arrow, Kenneth J and Dasgupta, Partha and Goulder, Lawrence H and Mumford, Kevin J and Oleson, Kirsten},
  journal={Environment and Development Economics},
  volume={18},
  number={4},
  pages={504--516},
  year={2013},
  publisher={Cambridge University Press}
}

@article{arrow2012sustainability,
  title={Sustainability and the measurement of wealth},
  author={Arrow, Kenneth J and Dasgupta, Partha and Goulder, Lawrence H and Mumford, Kevin J and Oleson, Kirsten},
  journal={Environment and development economics},
  volume={17},
  number={3},
  pages={317--353},
  year={2012},
  publisher={Cambridge University Press}
}

@article{morgan2025human,
  title={Human culture is uniquely open-ended rather than uniquely cumulative},
  author={Morgan, Thomas JH and Feldman, Marcus W},
  journal={Nature Human Behaviour},
  volume={9},
  number={1},
  pages={28--42},
  year={2025},
  publisher={Nature Publishing Group UK London}
}

@article{corominas2018zipf,
  title={Zipf’s Law, unbounded complexity and open-ended evolution},
  author={Corominas-Murtra, Bernat and Seoane, Lu{\'\i}s F and Sol{\'e}, Ricard},
  journal={Journal of the Royal Society Interface},
  volume={15},
  number={149},
  pages={20180395},
  year={2018},
  publisher={The Royal Society}
}

@article{adams2017formal,
  title={Formal definitions of unbounded evolution and innovation reveal universal mechanisms for open-ended evolution in dynamical systems},
  author={Adams, Alyssa and Zenil, Hector and Davies, Paul CW and Walker, Sara Imari},
  journal={Scientific reports},
  volume={7},
  number={1},
  pages={997},
  year={2017},
  publisher={Nature Publishing Group UK London}
}

@article{ruiz2008enabling,
  title={Enabling conditions for ‘open-ended evolution’},
  author={Ruiz-Mirazo, Kepa and Umerez, Jon and Moreno, Alvaro},
  journal={Biology \& Philosophy},
  volume={23},
  pages={67--85},
  year={2008},
  publisher={Springer}
}

@book{clark2010mathematical,
  title={Mathematical bioeconomics: the mathematics of conservation},
  author={Clark, Colin W},
  year={2010},
  publisher={John Wiley \& Sons}
}

@book{barro2025economic,
  title={Economic growth},
  author={Barro, Robert J and Sala-i-Martin, Xavier I},
  year={2025},
  publisher={MIT press}
}

@article{scheffer2009early,
  title={Early-warning signals for critical transitions},
  author={Scheffer, Marten and Bascompte, Jordi and Brock, William A and Brovkin, Victor and Carpenter, Stephen R and Dakos, Vasilis and Held, Hermann and Van Nes, Egbert H and Rietkerk, Max and Sugihara, George},
  journal={Nature},
  volume={461},
  number={7260},
  pages={53--59},
  year={2009},
  publisher={Nature Publishing Group}
}

@article{scheffer2001catastrophic,
  title={Catastrophic shifts in ecosystems},
  author={Scheffer, Marten and Carpenter, Steve and Foley, Jonathan A and Folke, Carl and Walker, Brian},
  journal={Nature},
  volume={413},
  number={6856},
  pages={591--596},
  year={2001},
  publisher={Nature Publishing Group UK London}
}

@book{may1973stability,
  title={Stability and Complexity in Model Ecosystems},
  author={May, Robert M.},
  year={1973},
  publisher={Princeton University Press},
  address={Princeton, NJ},
  isbn={978-0691081304},
  series={Monographs in Population Biology}
}

@article{hochberg2017innovation,
  title={Innovation: an emerging focus from cells to societies},
  author={Hochberg, Michael E and Marquet, Pablo A and Boyd, Robert and Wagner, Andreas},
  journal={Philosophical Transactions of the Royal Society B: Biological Sciences},
  volume={372},
  number={1735},
  pages={20160414},
  year={2017},
  publisher={The Royal Society}
}

@article{rhode2005world,
  title={The World Economy: Historical Statistics. By Angus Maddison. Paris: OECD, 2003. Pp. 384. 24.},
  author={Rhode, Paul W},
  journal={The Journal of Economic History},
  volume={65},
  number={1},
  pages={283--284},
  year={2005},
  publisher={Cambridge University Press}
}

@article{weinberger2017innovation,
  title={Innovation and the growth of human population},
  author={Weinberger, Vanessa P and Qui{\~n}inao, Crist{\'o}bal and Marquet, Pablo A},
  journal={Philosophical Transactions of the Royal Society B: Biological Sciences},
  volume={372},
  number={1735},
  pages={20160415},
  year={2017},
  publisher={The Royal Society}
}

@book{schumpeter1942capitalism,
  title     = {Capitalism, Socialism and Democracy},
  author    = {Schumpeter, Joseph A.},
  year      = {1942},
  publisher = {Harper \& Brothers},
  address   = {New York}
}

@article{mokyr2015history,
  title={The history of technological anxiety and the future of economic growth: Is this time different?},
  author={Mokyr, Joel and Vickers, Chris and Ziebarth, Nicolas L},
  journal={Journal of economic perspectives},
  volume={29},
  number={3},
  pages={31--50},
  year={2015},
  publisher={American Economic Association 2014 Broadway, Suite 305, Nashville, TN 37203-2418}
}

@book{daly1997beyond,
  title={Beyond growth: the economics of sustainable development},
  author={Daly, Herman E},
  year={1997},
  publisher={Beacon press}
}

@article{romer1990endogenous,
  title={Endogenous technological change},
  author={Romer, Paul M},
  journal={Journal of political Economy},
  volume={98},
  number={5, Part 2},
  pages={S71--S102},
  year={1990},
  publisher={The University of Chicago Press}
}

@article{aghion1997endogenous,
  title={Endogenous Growth Theory},
  author={Aghion, Philippe and Howitt, Peter},
  journal={MIT Press Books},
  volume={1},
  year={1997},
  publisher={The MIT Press}
}

@article{bardoscia2017statistical,
  title={Statistical mechanics of complex economies},
  author={Bardoscia, Marco and Livan, Giacomo and Marsili, Matteo},
  journal={Journal of Statistical Mechanics: Theory and Experiment},
  volume={2017},
  number={4},
  pages={043401},
  year={2017},
  publisher={IOP Publishing}
}

@book{cardy2012finite,
  title={Finite-size scaling},
  author={Cardy, John},
  volume={2},
  year={2012},
  publisher={Elsevier}
}

@article{galor2000population,
  title={Population, technology, and growth: From Malthusian stagnation to the demographic transition and beyond},
  author={Galor, Oded and Weil, David N},
  journal={American economic review},
  volume={90},
  number={4},
  pages={806--828},
  year={2000},
  publisher={American Economic Association}
}

@article{barro1989fertility,
  title={Fertility choice in a model of economic growth},
  author={Barro, Robert J and Becker, Gary S},
  journal={Econometrica: journal of the Econometric Society},
  pages={481--501},
  year={1989},
  publisher={JSTOR}
}

@article{dasgupta2002confronting,
  title={Confronting the environmental Kuznets curve},
  author={Dasgupta, Susmita and Laplante, Benoit and Wang, Hua and Wheeler, David},
  journal={Journal of economic perspectives},
  volume={16},
  number={1},
  pages={147--168},
  year={2002},
  publisher={American Economic Association}
}

@article{stern2004rise,
  title={The rise and fall of the environmental Kuznets curve},
  author={Stern, David I},
  journal={World development},
  volume={32},
  number={8},
  pages={1419--1439},
  year={2004},
  publisher={Elsevier}
}

@article{grossman1995economic,
    author = {Grossman, Gene M. and Krueger, Alan B.},
    title = {Economic Growth and the Environment*},
    journal = {The Quarterly Journal of Economics},
    volume = {110},
    number = {2},
    pages = {353-377},
    year = {1995},
    month = {05},
    issn = {0033-5533},
}

@book{ostrom1990governing,
  title={Governing the commons: The evolution of institutions for collective action},
  author={Ostrom, Elinor},
  year={1990},
  publisher={Cambridge university press}
}

@article{Fanning2025,
   author = {Andrew L. Fanning and Kate Raworth},
   issue = {8083},
   journal = {Nature},
   month = {10},
   pages = {47-56},
   publisher = {Nature Publishing Group},
   title = {Doughnut of social and planetary boundaries monitors a world out of balance},
   volume = {646},
   year = {2025}
}

@article{raworth2017doughnut,
  title={A Doughnut for the Anthropocene: humanity's compass in the 21st century},
  author={Raworth, Kate},
  journal={The lancet planetary health},
  volume={1},
  number={2},
  pages={e48--e49},
  year={2017},
  publisher={Elsevier}
}

@article{langridge2023ecological,
  title={An ecological basic income? Examining the ecological credentials of basic income through a review of selected pilot interventions},
  author={Langridge, Nicholas and Buchs, Milena and Howard, Neil},
  journal={Basic Income Studies},
  volume={18},
  number={1},
  pages={47--87},
  year={2023},
  publisher={De Gruyter}
}

@incollection{brock2013polluted,
  title={A polluted golden age},
  author={Brock, William A},
  booktitle={Economics of Natural \& Environmental Resources (Routledge Revivals)},
  pages={441--461},
  year={2013},
  publisher={Routledge}
}

@book{Malthus1803,
  author    = {Malthus, Thomas Robert},
  title     = {An Essay on the Principle of Population},
  edition   = {2},
  year      = {1803},
  publisher = {J. Johnson},
  address   = {London}
}

@article{solow1973end,
  title={Is the End of the World at Hand?},
  author={Solow, Robert M},
  journal={Challenge},
  volume={16},
  number={1},
  pages={39--50},
  year={1973},
  publisher={Taylor \& Francis}
}

@article{brander1998simple,
  title={The simple economics of Easter Island: A Ricardo-Malthus model of renewable resource use},
  author={Brander, James A and Taylor, M Scott},
  journal={American economic review},
  pages={119--138},
  year={1998},
  publisher={JSTOR}
}

@book{neumayer2013weak,
  title     = {Weak versus Strong Sustainability: Exploring the Limits of Two Opposing Paradigms},
  author    = {Neumayer, Eric},
  year      = {2013},
  edition   = {4},
  publisher = {Edward Elgar Publishing},
  address   = {Cheltenham, UK and Northampton, MA}
}

@article{graedel2015materials,
  title={On the materials basis of modern society},
  author={Graedel, Thomas E and Harper, Ermelina M and Nassar, Nedal T and Reck, Barbara K},
  journal={Proceedings of the National Academy of Sciences},
  volume={112},
  number={20},
  pages={6295--6300},
  year={2015},
  publisher={National Academy of Sciences}
}

@article{weisz2015industrial,
  title={Industrial Ecology: The role of manufactured capital in sustainability},
  author={Weisz, Helga and Suh, Sangwon and Graedel, Thomas E},
  journal={Proceedings of the National Academy of Sciences},
  volume={112},
  number={20},
  pages={6260--6264},
  year={2015},
  publisher={National Academy of Sciences}
}

@article{heron2025anti,
  title={For an Anti-Imperialist Ecological Modernity},
  author={Heron, Kai and Pedregal, Alejandro and Luki{\'c}, Nemanja},
  journal={Journal of Labor and Society},
  volume={1},
  number={aop},
  pages={1--55},
  year={2025},
  publisher={Brill}
}

@misc{asafuAdjaye2015ecomodernist,
  title        = {An Ecomodernist Manifesto},
  author       = {Asafu-Adjaye, John and Blomqvist, Linus and Brand, Stewart and Brook, Barry and DeFries, Ruth and Ellis, Erle and Foreman, Christopher and Keith, David and Lewis, Martin and Lynas, Mark and Nordhaus, Ted and Pielke Jr., Roger and Pritzker, Rachel and Roy, Joyashree and Sagoff, Mark and Shellenberger, Michael and Stone, Robert and Wigley, Tom},
  year         = {2015},
  howpublished = {\url{http://www.ecomodernism.org/manifesto}},
  note         = {Accessed 2015}
}

\pagebreak
\appendix
\counterwithin{figure}{section}
\renewcommand{\thefigure}{S\arabic{figure}}
\renewcommand\theequation{A\arabic{equation}}
\setcounter{equation}{0}

\section{Supporting Information}

Here we provide a detailed derivation of the main calculations of the model, complementary to the Materials and Methods section of the paper. We also perform an asymptotic analysis that precisely characterizes the phase transition, and we present a generalization of the model to account for more demanding innovation in a pristine environment.

We start from the dynamic equations Eqs.~\eqref{eq:lambda},~\eqref{eq:epsilon} and \eqref{eq:eta} in terms of the rescaled variables $\lambda_i,\epsilon$ and $\eta$ defined in Eq.~\eqref{eq:def:resc_vars} and parameters $\theta_i,\gamma_i,\delta_i$ and $\sigma_i$. For convenience we reproduce these equations here:
\begin{eqnarray}
\frac{d \lambda_i}{d s} &=&c\max\left\{\theta_i(\eta-\Lambda)-\gamma_i(1-\epsilon)-\lambda_i,0\right\} \,,\qquad \Lambda = \sum_{i=1}^I \lambda_i
\label{eq:SI_tech} \\
\frac{\tau_E}{\tau_L}\frac{d\epsilon}{ds}&=& \epsilon\left(1-\epsilon\right)-\sum_i \delta_i \epsilon ^ u \lambda_i \epsilon \label{eq:SI_env} \\
\frac{\tau_N}{\tau_L}\frac{d \eta}{ds}&=& \eta\left(1+\sum_{i=1}^I \sigma_i \lambda_i-\frac{\eta}{\eta_0}\right)\,,\qquad\eta_0=\frac{N_0}{E_0}
\label{eq:SI_pop}
\end{eqnarray}
where $t = s \tau_L$ is the rescales time, with respect to the labor timescale $\tau_L$, and $u=0$ or $1$ depending on whether the extraction rate is constant or proportional to $E/E_0=\epsilon$.

\subsection{Details of the derivation}

We will examine the equilibrium of the system, setting all derivatives of equations (\ref{eq:SI_tech}), (\ref{eq:SI_env}), and (\ref{eq:SI_pop}) to zero. For $\lambda_i$, we get  
\begin{equation}
    \lambda_i=\theta_i(\eta-\Lambda)-\gamma_i(1-\epsilon),
\end{equation}
where
\begin{equation}
    \Lambda=\frac{\Theta\eta}{1+\Theta}-\frac{\Gamma(1-\epsilon)}{1+\Theta}
\end{equation}
is the total scale of production, with $\Theta=\sum_{i}\theta_i$ and $\Gamma=\sum_{i}\gamma_i$. 
The condition for a new technology to take off is that, in its absence, $\dot \lambda_i>0$. This implies $\gamma_i\le\kappa\theta_i$ with the threshold 
\begin{equation}
\kappa=\frac{\eta-\Lambda}{1-\epsilon}=\frac{1}{1+\Theta}\left(\frac{\eta}{1-\epsilon}+\Gamma\right). 
\label{eq:SI_kappa}
\end{equation}

To obtain the equivalent equation for the steady states of the population in terms of $\eta$, we write the equilibrium value for technology $i$ in terms of $\epsilon$ and $\eta$ alone,
\begin{equation}
    \lambda_i=\frac{\theta_i}{1+\Theta} \eta+\theta_i\left(\frac{\Gamma}{1+\Theta}-\frac{\gamma_i}{\theta_i}\right)(1-\epsilon),
\end{equation} 
and, assuming $\sigma_i=\theta_i/\alpha$, then we have
\begin{align*}
    \alpha\sum_{i}\sigma_i \lambda_i &=\sum_{i}\left(\frac{\theta_i^2}{1+\Theta}\eta+\theta_i^2\left(\frac{\Gamma}{1+\Theta}-\frac{\gamma_i}{\theta_i}\right)(1-\epsilon)\right) 
   \\
   &=\frac{\Psi}{1+\Theta}\eta + \left(\frac{\Psi\Gamma}{1+\Theta}-\Delta\right) (1-\epsilon),
\end{align*}
where $\Psi=\sum_{i}\theta_i^2$ and $\Delta=\sum_{i}\gamma_i\theta_i$. Using this relationship, we can rewrite the steady state for $\eta$ as
 \[\eta  = \eta_0 (1+\sum_{i=1}^I \sigma_i \lambda_i) = \eta_0+\frac{\eta_0}{\alpha}\frac{\Psi}{1+\Theta}\eta + \frac{\eta_0}{\alpha}\left(\frac{\Psi\Gamma}{1+\Theta}-\Delta\right) (1-\epsilon),
 \]
then
\begin{align}
    \eta  = \eta_0 \frac{1+\frac{\Sigma}{\alpha}(1-\epsilon)}{1-\eta_0\frac{\chi}{\alpha}},\qquad\text{with}\quad
\chi = \frac{\Psi}{1+\Theta},\quad \Sigma=\left(\frac{\Psi\Gamma}{1+\Theta}-\Delta\right).
\label{eq:SI_eta}
\end{align}

For the equation of the environmental variable $\epsilon$, we proceed similarly. If $\delta_i=\beta\gamma_i$, then 
\begin{align*}
   \frac{1}{\beta}\sum_{i}\delta_i \lambda_i &=\sum_i\left(\frac{\gamma_i\theta_i}{1+\Theta}\eta+\gamma_i\theta_i\left(\frac{\Gamma}{1+\Theta}-\frac{\gamma_i}{\theta_i}\right)(1-\epsilon)\right) 
   \\
   &=\frac{\Delta}{1+\Theta}\eta + \left(\frac{\Delta\Gamma}{1+\Theta}-\Phi\right) (1-\epsilon),
\end{align*}
with $\Phi=\sum_{i}\gamma_i^2$. We then analyze each case of the extraction rate, considering both the constant and proportional to $\epsilon$ forms. 

\subsubsection{Constant extraction rate, $u=0$}

In this case, the equilibrium for the environment implies
\begin{equation*}
    1-\epsilon=\sum_i \delta_i \lambda_i= \beta \frac{\Delta}{1+\Theta}\eta + \beta \left(\frac{\Delta\Gamma}{1+\Theta}-\Phi\right) (1-\epsilon) =\beta \frac{\Delta}{1+\Theta}\eta - \beta \Omega (1-\epsilon),
\end{equation*}
with $\Omega=\Phi-\frac{\Delta\Gamma}{1+\Theta}$. Then
\begin{equation*}
    \epsilon = 1 - \frac{\beta}{1+\beta \Omega}\frac{\Delta}{1+\Theta}\eta,
\end{equation*}
and inputting (\ref{eq:SI_eta}) we get
\begin{align}
    \epsilon = 1- \frac{\frac{\beta}{1+\beta \Omega}\frac{\Delta}{1+\Theta} \eta_0}{1-\frac{\chi}{\alpha} \eta_0-\frac{\Sigma}{\alpha}\frac{\beta}{1+\beta \Omega}\frac{\Delta}{1+\Theta} \eta_0 }
\end{align}
for $\eta_0$ constant. For the population, we get the equilibrium
\begin{align}
    \eta = \frac{\eta_0}{1-\frac{\chi}{\alpha} \eta_0-\frac{\Sigma}{\alpha}\frac{\beta}{1+\beta \Omega}\frac{\Delta}{1+\Theta}\eta_0}.
\end{align}
The threshold $\kappa$ reads simply
\begin{align}
    \kappa = \frac{1+\beta \Phi}{ \beta \Delta}.
\end{align}

\subsubsection{Extraction rate proportional to $\epsilon$, $u=1$}

In this case, the equilibrium for $\epsilon$ implies
\begin{equation*}
    1-\epsilon=\epsilon \sum_i \delta_i \lambda_i= \epsilon \beta \left( \frac{\Delta}{1+\Theta}\eta - \Omega (1-\epsilon)\right).
\end{equation*}
By using~\eqref{eq:SI_eta}, it follows that
    \begin{align*}
   \frac{1}{\epsilon} &=1+\beta\frac{ \Delta}{1+\Theta}\eta_0 \frac{1+\frac{\Sigma}{\alpha}(1-\epsilon)}{1-\eta_0\frac{\chi}{\alpha}} - \beta\Omega (1-\epsilon)
   \\
   &=1+\frac{\beta \Delta}{1+\Theta}\frac{\eta_0}{1-\eta_0\frac{\chi}{\alpha}}+ \beta (1-\epsilon)\left(\frac{\Sigma\Delta\eta_0}{\alpha(1+\Theta)(1-\eta_0\frac{\chi}{\alpha})}-\Omega\right)
   \\
   &=1+A+B(1-\epsilon),
  \end{align*}
  where
  \begin{equation}
    A=\frac{\beta\Delta\eta_0}{(1+\Theta)(1-\eta_0\frac{\chi}{\alpha})}\,,\qquad B=\beta\left[\frac{\Delta\Sigma \eta_0}{\alpha(1+\Theta)(1-\eta_0\frac{\chi}{\alpha})}-\Omega\right].
\end{equation}
The solution for $\epsilon$ is 
\begin{equation}
\label{eq:constN0}
    \epsilon=\frac{1+A+B\pm\sqrt{(1+A+B)^2-4B}}{2B}
\end{equation}
The coefficient $A$ diverges when $\eta_0\frac{\chi}{\alpha}\to 1$, which marks the instability transition at $\alpha_u=\eta_0 \chi$. 

It is also possible to solve the stationary equations for the case where the carrying capacity $\eta_0$ of the population depends on $\epsilon$, in particular when $\eta_0(\epsilon)=\eta_{00}(1-\epsilon)$. It can be shown that, also in this case, the system exhibits the same sequence of phase transitions discussed for the case where $\eta_0$ is a constant.


\subsection{Asymptotic Analysis of the Model and Phase transition ($u=1$)} 

\label{app:analytic} 
When the number $I_{act}$ of technologies is large, the quantities $\Theta,\Gamma,\ldots$ can be approximated by expectations over the random variables $\theta$ and $\gamma$, conditional on the constraint $\gamma\le\kappa\theta$. In particular,
    \begin{align*}
    \frac{1}{I_{act}}\sum_{i}\theta^x\gamma^y = \mathbb E\left[\theta^x\gamma^y|\gamma\leq\kappa\theta\right] = \frac{\mathbb{E}[\theta^x \gamma^y \cdot \mathbf{1}_{\{\gamma \le \kappa\theta\}}]}{\mathbb{P}(\gamma \le \kappa\theta)},
    \end{align*}
where $\theta$ and $\gamma$ are independent exponential random variables with means $\bar \theta$ and $\bar \gamma$, respectively. The joint distribution thus factorizes as $ p(\theta, \gamma) = p_{\bar \theta}(\theta) \cdot p_{\bar \gamma}(\gamma)$, and the conditional expectation takes the form
\begin{align*}
\mathbb{E}\left[\theta^x\gamma^y \mid \gamma \le \kappa\theta\right] =
\frac{\displaystyle\int_0^\infty \left( \int_0^{\kappa\theta} \theta^x \gamma^y \, p_\gamma(\gamma) \, d\gamma \right) p_\theta(\theta) \, d\theta}{\displaystyle\int_0^\infty \left( \int_0^{\kappa\theta} p_\gamma(\gamma) \, d\gamma \right) p_\theta(\theta) \, d\theta}.
\end{align*}
With this general formula, we will compute all quantities required to study the $I\rightarrow\infty$ limit. First notice that
 \begin{align*}
 \mathbb{E}[\mathbf{1}_{\{\gamma \le \kappa\theta\}}] &= 
 \frac{1}{\bar\theta}\int_0^\infty d\theta\,\, e^{-\theta/\bar\theta}\left(1-e^{-\kappa\theta/\bar\gamma}\right)  
 \\
 &= 
\frac{\kappa\bar\theta}{\bar\gamma+\kappa\bar\theta}=\frac{z}{1+z},
 \end{align*}
where we have defined $z=\kappa\bar\theta/\bar\gamma = \kappa/\bar\kappa$. By some similar calculations, the following analytical expressions hold:
\begin{align*}
    \frac{\Theta}{I_{act}} =\frac{\bar\theta  (2 \bar\gamma +\bar\theta  \kappa )}{\bar\gamma +\bar\theta  \kappa } = \bar\theta\,\frac{ 2+z}{1+z},   \qquad
   \frac{\Gamma}{I_{act}} = \frac{\bar\gamma  \bar\theta  \kappa }{\bar\gamma +\bar\theta  \kappa }=\bar\gamma\frac{z}{1+z}, \qquad
    \frac{\Delta}{I_{act}}=\frac{\bar\gamma  \bar\theta ^2 \kappa  (3 \bar\gamma +\bar\theta  \kappa )}{(\bar\gamma +\bar\theta  \kappa )^2}=\bar\gamma\bar\theta\frac{z(3+z)}{(1+z)^2}
    \\
 \\
    \frac{\Psi}{I_{act}}=\frac{2 \bar\theta  (\bar\gamma +\bar\theta  \kappa ) }{\kappa }\left(1-\frac{\bar\gamma ^3}{(\bar\gamma +\bar\theta  \kappa )^3}\right)=2\bar\theta^2\frac{3+3z+z^2}{(1+z)^2},\qquad
    \frac{\Phi}{I_{act}}=\frac{2 \bar\gamma ^2 \bar\theta ^2 \kappa ^2}{(\bar\gamma +\bar\theta  \kappa )^2} = \bar\gamma^2\frac{2z^2}{(1+z)^2}.
\end{align*}

The auxiliary quantities can also be expressed in terms of $z$, analytically in the large $I_{act}$ limit as combinations of the moments above:
\[
    \chi = \frac{\Psi}{1+\Theta}\simeq 2 \bar\theta  \left(\frac{\bar\gamma ^2}{(\bar\gamma +\bar\theta  \kappa ) (2 \bar\gamma +\bar\theta  \kappa )}+1\right) = \bar\theta\frac{2(3+3z+z^2)}{(1+z)(2+z)},
    \]
   in a similar way
   \[
    \frac{\Sigma}{I_{act}} =\left(\frac{\Psi\Gamma}{I_{act}(1+\Theta)}-\frac{\Delta}{I_{act}}\right) \simeq \frac{\bar\gamma  \bar\theta ^3 \kappa ^2}{(\bar\gamma +\bar\theta  \kappa ) (2 \bar\gamma +\bar\theta  \kappa )} = \bar\gamma\bar\theta\frac{z^2}{(1+z)(2+z)},
    \]
    and
    \[
    \frac{\Omega}{I_{act}} = \frac{\Phi}{I_{act}}-\frac{\Delta\Gamma}{I_{act}(1+\Theta)} \simeq  \frac{\bar\gamma ^2 \theta ^2 \kappa ^2}{(\bar\gamma +\bar\theta  \kappa ) (2 \bar\gamma +\bar\theta  \kappa )}=\bar\gamma^2\frac{z^2}{(1+z)(2+z)}.
\]
   The two terms $A$ and $B$ are given by
   \begin{align*}
   A = \frac{\beta \eta_0}{(1-\eta_0\frac{\chi}{\alpha})}\frac{\Delta}{(1+\Theta)}&\simeq \frac{\beta\eta_0\bar\gamma}{1-\frac{\eta_0}{\alpha}\bar\theta\,\frac{2(z^2+3z+3)}{(1+z)(2+z)}}\,\frac{z(3+z)}{(1+z)(2+z)}
   \\
   &=\frac{\beta\eta_0\bar\gamma\,z(3+z)}{(1+z)(2+z)-2\bar\theta\frac{\eta_0}{\alpha}(z^2+3z+3)},
   \end{align*}
   and
   \begin{align*}
   B &=\beta\left[\frac{\Delta\Sigma \eta_0}{\alpha(1+\Theta)(1-\eta_0\frac{\chi}{\alpha})}-\Omega\right]
   \\
   &\simeq \beta\left[\frac{\eta_0\bar\gamma\,z(3+z)}{\alpha (1+z)(2+z)-2\eta_0\bar\theta(z^2+3z+3)}\bar\gamma\bar\theta\frac{z^2}{(1+z)(2+z)}-\bar\gamma^2\frac{z^2}{(1+z)(2+z)}\right] I_{act}
    \\
   &\simeq \beta\bar\gamma^2I_{act}\frac{z^2}{(1+z)(2+z)}\left[\frac{\eta_0\bar\theta\,z(3+z)}{\alpha (1+z)(2+z)-2\eta_0\bar\theta(z^2+3z+3)}-1\right] 
   \\
   &=\beta \bar\gamma^2I_{act}(3\eta_0\bar\theta-\alpha)\frac{z^2}{\alpha(1+z)(2+z)-2\eta_0\bar\theta(z^2+3z+3)}
   \end{align*}

The solution for the environment, given by Eq.~(\ref{eq:constN0}), follows directly by using the inputs $A$ and $B$. Combined to Eqs. (\ref{eq:eta}) and (\ref{eq:kappa}), we obtain a transcendental equation for $\kappa$, which can be solved numerically. The resulting values of $\kappa$ are consistent with those obtained from simulations (Fig. \ref{fig:analytic}).

To characterize the phase transition, we analyze the asymptotic behavior of the system. In the limit of small $\alpha$, $\kappa$ diverges and $\epsilon \approx 0$, whereas for large $\alpha$, $\kappa \approx 0$ and $\epsilon \approx 1$. 

\subsubsection*{The large $\alpha$ regime}

For large $\alpha$ we look for a solution to the equations where the number $I_{act}$ of active technologies  diverges and $\kappa\to 0$. So we'll consider the leading orders in $\kappa$ and $1/I_{act}$. To leading order
\begin{equation}\nonumber
    A\simeq a\kappa+\ldots,\quad B\simeq-b\kappa^2 I_{act}(1-b_1\kappa)+\ldots,\quad a=\frac{3\beta\eta_0\bar\theta}{2(1-3\frac{\eta_0}{\alpha} \bar{\theta} )},\quad b=\frac{\beta\bar\theta^2}{2},\quad b_1=\frac{3\bar\theta}{2\bar\gamma}\frac{\alpha-2\eta_0\bar\theta}{\alpha-3\eta_0\bar\theta }.
\end{equation}
However, we want to analyze the conjoint effect. To do so, we assume that there is some constant $C\neq0$ such that $I_{act}\kappa^2\rightarrow C$ when $\kappa\rightarrow0$ and $I_{act}\rightarrow\infty$. As a consequence to the first order $B\simeq -Cb(1-b_1\kappa)$. In this case, the equation for the steady state of the environment becomes
\begin{align*}
\epsilon &\simeq\frac{1+a\kappa-Cb(1-b_1\kappa)-\sqrt{(1+a\kappa-Cb(1-b_1\kappa))^2+4Cb(1-b_1\kappa)}}{-2Cb(1-b_1\kappa)}
\\
&=1-\xi\kappa+O(\kappa^2),\qquad \xi:=\frac{a}{1+bC},
\end{align*}
which confirms that $\epsilon\to 1$ as $\kappa\to 0$. Using equation~\eqref{eq:SI_eta}, we have that for $\kappa$ small
\[
\eta\simeq\frac{\eta_0}{1-\frac{\eta_0}{\alpha}\chi}+\frac{\eta_0\Sigma\xi \kappa}{\alpha(1-\frac{\eta_0}{\alpha}\chi)},
\]
and then the equation for $\kappa$ reads
\begin{eqnarray*}
    \kappa&=&\frac{1}{1+\Theta}\left(\frac{\eta}{1-\epsilon}+\Gamma\right)
\simeq 
\frac{1}{I_{act}\bar\theta\,\frac{ 2+z}{1+z}}
\left(\frac{\eta_0}{\xi\kappa(1-\frac{\eta_0}{\alpha}\chi)}+
\frac{\eta_0\Sigma}{\alpha(1-\frac{\eta_0}{\alpha}\chi)}+
\frac{\bar\gamma z}{1+z}I_{act}\right)=:\mathcal T_1+\mathcal T_2+\mathcal T_3.
\end{eqnarray*}
We treat each term separately:
\[
\mathcal T_1 = \frac{\eta_0(1+z)}{\xi I_{act}\kappa\bar\theta (2+z)(1-\frac{\eta_0}{\alpha}\chi)}=\frac{\eta_0}{\xi I_{act}\kappa\bar\theta (1-\frac{\eta_0}{\alpha}\chi)}\frac{\bar\gamma+\bar\theta\kappa}{2\bar\gamma+\bar\theta\kappa}
\]
and
\[
\mathcal T_2 = \frac{ \eta_0\bar\gamma}{\alpha(1-\frac{\eta_0}{\alpha}\chi)}\frac{z^2}{(2+z)^2}=\frac{ \eta_0\bar\gamma\bar\theta^2}{\alpha(1-\frac{\eta_0}{\alpha}\chi)}\frac{\kappa^2}{(2\bar\gamma+\bar\theta\kappa)^2}
,\qquad
\mathcal T_3 
=\frac{\bar\gamma\kappa}{2\bar\gamma+\bar\theta\kappa}.
\]
In the limit $\alpha\rightarrow \infty$, the equation for $\kappa$ becomes
\[
\kappa = \frac{\eta_0}{\xi I_{act}\kappa\bar\theta}\frac{\bar\gamma+\bar\theta\kappa}{2\bar\gamma+\bar\theta\kappa}+\frac{\bar\gamma\kappa}{2\bar\gamma+\bar\theta\kappa}=\sqrt{\frac{\eta_0}{I_{act}\xi\bar\theta }},
\]
which is consistent with the hypothesis $I_{act}\kappa^2\rightarrow C$. Moreover we get an explicit expression for the constant $C$:
\[
I_{act}\kappa^2\sim C=\frac{\eta_0}{\xi\bar\theta }=\frac{\eta_0}{\bar\theta}\frac{1+bC}{a}=\frac{\eta_0}{a\bar\theta-b\eta_0}=\frac{1}{\beta\bar\theta^2}.
\]
By continuity, this solution shifts to the right for large values of $\alpha$. Yet, the approximations employed used are not appropriate for the solution with finite $\kappa$, that will be discussed below. Taking into account that 
\[
I_{act} = \sum_{i=1}^I\mathbb P[\gamma_i<\kappa\theta_i] = I\frac{\kappa\bar\theta}{\bar\gamma+\kappa\bar\theta},
\]
which means that $I_{act}\sim \kappa I$ if $\kappa$ is small, this also implies the following scaling relations as $T\to \infty$ in the large $\alpha$ regime:
\begin{equation}
\kappa\sim I^{-1/3}\,,\qquad I_{act}\sim I^{2/3} \,,\qquad N\sim I^0,\qquad E_0-E\sim I^{-1/3}
\end{equation}
The convergence of $E$ to $E_0$ for large $\alpha$ is non-trivial, because it requires that
\begin{equation}
    \sum_i\delta_i s_i=\frac{\Delta}{1+\Theta}N-\Omega(E_0-E)
\end{equation}
vanishes as $I\to\infty$. This is consistent with a solution where $\kappa\to 0$ in this limit, because in the first term $\Delta/(1+\Theta)\sim \kappa$ and in the second $\Omega\sim \kappa^2 I_{act}$.

\subsubsection*{The small $\alpha$ regime}

\begin{figure}
    \centering
    \includegraphics[width=\linewidth]{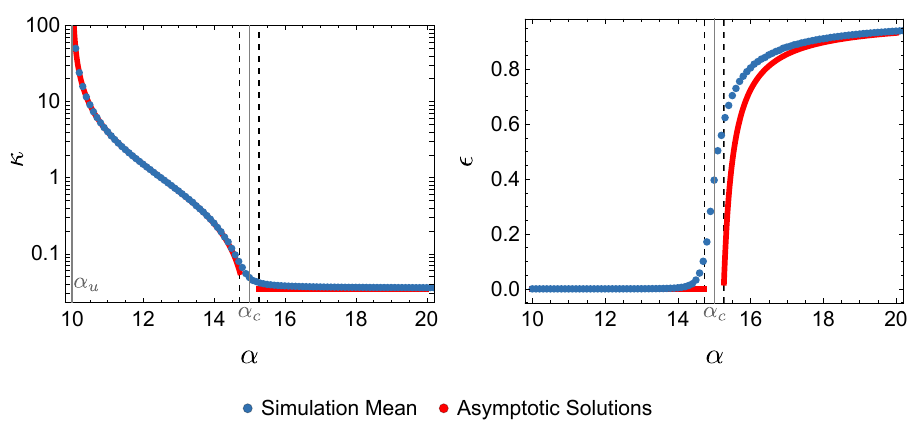}
    \caption{Comparison of asymptotic solutions and simulation results for $\kappa$ and $\epsilon$ as functions of $\alpha$, with $T=10^6$.  All other parameters are consistent with those used in the figures in the main text. The region between dashed lines corresponds to the critical region, with width $\alpha_c \pm \delta \alpha$, with $\delta \alpha \sim T^{-\nu}$.}
    \label{fig:analytic}
\end{figure}

In the limit $I\to\infty$, we obtain from Eqs.~\eqref{eq:SI_kappa} and~\eqref{eq:SI_eta} the following expression:
\begin{align*}
    \kappa \simeq\frac{1}{\Theta}\left(\frac{\eta_0\Sigma }{\alpha(1-\frac{\eta_0}{\alpha}\chi)}+\Gamma\right) = 
    \frac{ \kappa  ((1-3\frac{\eta_0}{\alpha} \bar\theta)+ z (1-\frac{\eta_0}{\alpha}\bar\theta))}{2 (1-3 \frac{\eta_0}{\alpha} \bar\theta  )+3 z (1-2 \frac{\eta_0}{\alpha} \bar\theta )+z^2 (1-2 \frac{\eta_0}{\alpha} \bar\theta )}
\end{align*}
which has three solutions: $\kappa= 0$, $\kappa=-\bar\kappa$ (neglected due to lack of physical meaning), and $\kappa = \bar\kappa\frac{ (\alpha -3 \eta_0 \bar\theta )}{2 \eta_0 \bar \theta -\alpha}$. 
Since we are interested in finite and positive values of $\kappa$, we examine the conditions under which the third solution is physically meaningful. For any given $\bar\theta$, this condition implies that $2 \eta_0\bar\theta<\alpha<3\eta_0\bar\theta$. Indeed, for the parameter value $\bar\theta = 0.5$, this yields $\alpha \leq \alpha_c\equiv 3\eta_0/2$, which corresponds to the critical point marking the phase transition in the system. Moreover, as $\alpha \rightarrow \eta_0$, the third solution diverges to infinity as in Figure~\ref{fig:analytic}. It is worth noting that the solution $\kappa = 0$ corresponds to an extrapolation in the limit $I\to\infty$.

\subsubsection*{The critical region}

For finite $I$, the system is expected to cross over from the $\alpha>\alpha_c$ regime, where $\kappa\sim T^{-1/3}$, to the $\alpha<\alpha_c$ regime, where $\kappa$ is finite. 
Within finite-size scaling theory~\cite{cardy2012finite}, this crossover can be described by the scaling ansatz:
\begin{equation*}
    \kappa=I^{-1/3}F\left((\alpha_c-\alpha)I^\nu\right),
\end{equation*}
which assumes that the critical point $\alpha_c$ at finite $I$ broadens into a region of width $I^{-\nu}$. 
The behavior $\kappa\sim I^{-1/3}$ for $\alpha>\alpha_c$ is recovered by assuming that $F(x)$ converges to a constant, $F(x)\to F_+$, when $x\to \infty$. 
Conversely, the $\alpha<\alpha_c$ regime requires that $F(x)\sim x^{1/(3\nu)}$ to recover a finite value of $\kappa$ independent on $I$.
This implies $\kappa\simeq A(\alpha_c-\alpha)^{1/(3\nu)}$ for $I\to\infty$, recovering the linear dependence predicted by our theory when $\nu=1/3$.

\end{document}